\documentclass[10pt,emptycopyrightspace]{ewsn-proc}

\usepackage[square,sort&compress]{natbib}
\setcitestyle{numbers}
\usepackage{amsmath}
\usepackage{color}
\usepackage{xspace}

\usepackage[british]{babel}
\usepackage[nolist]{acronym}
\usepackage[binary-units]{siunitx}
\DeclareSIUnit\dBm{\decibel{}m}
\DeclareSIUnit\dBi{\decibel{}i}

\usepackage{graphicx}
\graphicspath{{images/}}
\usepackage{balance}
\usepackage[breaklinks]{hyperref}
\hypersetup{
    pdftitle = {Mitigating Inter-network Interference in LoRa Low-Power Wide-Area Networks},
    pdfauthor = {Thiemo Voigt, Martin Bor, Utz Roedig, Juan M. Alonso}
}

\newcommand{\eg}{e.\,g.\/~}
\newcommand{\ie}{i.\,e.,\/~}

\newcommand{\simulator}{LoRaSim\xspace}

\newcommand{\fakeparagraph}[1]{\vspace{0mm}\noindent\textbf{#1.}}
\newcommand{\spida}      {SPIDA\xspace}

\begin{document}






\title{Mitigating Inter-network Interference in LoRa Networks}

\numberofauthors{3}

\author{
%
\alignauthor Thiemo Voigt \\
    \affaddr{Uppsala University, Sweden}\\
    \affaddr{SICS Swedish ICT}\\
    \email{thiemo@sics.se}
\alignauthor Martin Bor, Utz Roedig \\
    \affaddr{Lancaster University, UK}\\
    \email{m.bor@lancaster.ac.uk\\u.roedig@lancaster.ac.uk}
\alignauthor Juan Alonso \\
    \affaddr{Univ.\@ Nac.\@ de Cuyo and Univ.\@ Nac.\@ de San Luis, Argentina}\\
    \email{jmalonso@uncu.edu.ar}
}



\maketitle

\begin{acronym}[]
\acro{WSN}{Wireless Sensor Networks}
\acro{LPWAN}{Low-Power Wide-Area Network}
\acro{LoRa}{Long Range}
\acro{LoRaWAN}{Long Range Wide Area Network}
\acro{MAC}{Medium Access Control}
\acro{IoT}{Internet of Things}
\acro{CCA}{Clear Channel Assessment}
\acro{CAD}{Carrier Activity Detection}
\acro{CSS}{Chirp Spread Spectrum}
\acro{FEC}{Forward Error Correction}
\acro{CF}{Carrier Frequency}
\acro{SF}{Spreading Factor}
\acro{TP}{Transmission Power}
\acro{SNR}{Signal to Noise Ratio}
\acro{BW}{Bandwidth}
\acro{CR}{Coding Rate}
\acro{CRC}{Cyclic Redundancy Check}
\acro{CDMA}{Code Division Multiple Access}
\acro{RSSI}{Received Signal Strength Indicator}
\acro{ACK}{Acknowledgement}
\acro{SRD}{Short Range Device}
\acro{ETSI}{European Telecommunications Standards Institute}
\acro{CEPT}{European Conference of Postal and Telecommunications Administrations}
\acro{FCC}{Federal Communications Commission}
\acro{CFR}{Code of Federal Regulations}
\acro{LBT}{Listen Before Talk}
\acro{AFA}{Adaptive Frequency Agility}
\acro{ERP}{Effective Radiated Power}
\acro{PER}{Packet Error Rate}
\acro{TCXO}{temperature compensated crystal oscillator}
\acro{DER}{Data Extraction Rate} 
\acro{NEC}{Network Energy Consumption}
\acro{FHSS}{Frequency Hopping Spread Spectrum}
\acro{DSSS}{Direct-Sequence Spread Spectrum}
\acro{RF}{Radio Frequency}
\end{acronym}

\begin{abstract}
\acf{LoRa} is a popular technology used to construct \acf{LPWAN} networks. Given the popularity of \ac{LoRa} it is likely that multiple independent \ac{LoRa}  networks are deployed in close proximity. In this situation, neighbouring networks interfere and methods have to be found to combat this interference.  In this paper we investigate the use of directional antennae and the use of multiple base stations as methods of dealing with inter-network interference. Directional antennae  increase signal strength at receivers without increasing transmission energy cost. Thus, the probability of successfully decoding the message in an interference situation is improved.  Multiple base stations can alternatively  be used to improve the probability of receiving a message in a noisy environment. We compare the effectiveness of these two approaches via simulation. Our findings show that both methods are able to improve \ac{LoRa} network performance in interference settings. However, the results show that the use of multiple base stations clearly outperforms the use of directional antennae. For example, in a setting where data is collected from 600 nodes which are interfered by four networks with 600 nodes each, using three base stations improves the \acf{DER} from 0.24 to 0.56 while the use of directional antennae provides an increase to only 0.32.

\end{abstract}



\section{Introduction}\label{sec:intro}

\acf{LoRa}  \acf{LPWAN} devices communicate directly with base stations which removes the need of constructing and maintaining a complex  multi-hop network. Multiple \ac{LoRa} networks may be deployed in the same physical space which leads to inter-network interference. For example, multiple smart city applications based on \ac{LoRa} may be deployed in the same area and interference between these networks will occur. To ensure acceptable network performance this inter-network interference must be managed appropriately. 

\ac{LoRa} transceivers can use orthogonal transmission settings (such as frequency, spreading factor, bandwidth) which in principle can be used to prevent inter-network interference. However, there are drawbacks which make this approach less viable in practical settings. First, transmitter settings have an impact on transmission properties such as range, reliability and energy consumption which prevents nodes to select parameters freely. Second, dynamically choosing parameters requires a complex protocol and network cooperation which current \ac{LoRa} systems do not support. Hence, current \ac{LoRa} deployments typically use a default static setting which leads to inter-network interference.  For these reasons it is desirable to find additional mechanisms for dealing with this interference. 

In this paper we focus on two practical alternative methods to deal with interference. Both methods aim at improving the chance of decoding a message in presence of interference. First, we consider directional antennae to improve signal strength at the receiver without increasing transmission energy cost. Second, we consider the use of multiple base stations to improve the probability of decoding a message at at least one receiver. While both methods have advantages and disadvantages in terms of practicality (such as cost, method of deployment, maintainability) it is the question which of the approaches is more effective. 

In this paper we answer this question by analysing the effectiveness of both approaches via comprehensive simulation. Our purpose built simulation environment is calibrated using \ac{LoRa} testbed experiments to ensure simulation results match as close as possible practical setups. Our findings show that both methods are able to improve \ac{LoRa} network performance in interference settings as it would be expected. However, the results demonstrate that the use of multiple base stations clearly outperforms the use of directional antennae. For example, in a setting where data is collected from 600 nodes which are interfered by four other \ac{LoRa} networks with 600 nodes each, the use of three base stations improves the \acf{DER} from 0.24 to 0.56 while the use of directional antennae increases it to 0.32.  

The main contributions of this paper are:
\begin{itemize}
\item We evaluate the impact of inter-network interference on \ac{LoRa} networks showing that such interference can drastically reduce the performance of a \ac{LoRa} network.
\item We quantify network performance gains by introducing directional antennae and multiple base stations.
\item We show that adding more base stations rather than equipping nodes with directional antennae is more efficient when mitigating \ac{LoRa} network interference.
\end{itemize}

In the next section, we present essential background on \ac{LoRa}. Section~\ref{sec:sim} describes briefly our previous work on \ac{LoRa}~\cite{bor16lora,bor2016lora} and the resulting simulation environment used for our evaluation. Section~\ref{sec:eval} describes the evaluation of performance gains by introducing directional antennae and multiple base stations. Before concluding, we discuss related work in Section~\ref{sec:related_work}.

\section{\acf{LoRa}}\label{sec:background}

\acf{LoRa} is a proprietary spread spectrum modulation technique by Semtech, derived from \ac{CSS}. Instead of modulating the message on a pseudorandom binary sequence, as is done in the well known \ac{DSSS}, LoRa uses a sweep tone that increases (upchirp) or decreases (downchirp) in frequency over time to encode the message. Spreading the signal over a wide bandwidth makes it less susceptible to noise and interference. \ac{CSS} in particular is resistant to Doppler effects (common in mobile applications) and multipath fading. A \ac{LoRa} receiver can decode transmissions \SI{20}{\dB} below the noise floor, making very long communication distances possible, while operating at a very low power. \ac{LoRa} transceivers available today can operate between \SIrange{137}{1020}{\MHz}, and therefore can also operate in licensed bands. However, they are often deployed in ISM bands (EU: \SI{868}{\MHz} and \SI{433}{\MHz}, USA: \SI{915}{\MHz} and \SI{433}{\MHz}). The \ac{LoRa} physical layer may be used with any MAC layer; however, \ac{LoRaWAN} is the currently proposed MAC. \ac{LoRaWAN} operates in a simple star topology.

A \ac{LoRa} transceiver has five runtime-adjustable transmission parameters: \ac{TP}, \ac{CF}, \ac{SF}, \ac{BW}, and \ac{CR}. These parameters have an influence on the transmission duration, energy consumption, robustness and range.

\fakeparagraph{\acf{TP}}
\ac{TP} on a LoRa receiver can be adjusted between \SI{-4}{\dBm} and \SI{20}{\dBm} in \SI{1}{\dB} steps. Because of regulatory and hardware limitations, however, this is often limited between \SI{2}{\dBm} and \SI{14}{\dBm}. \ac{TP} has a direct influence on energy consumption and the range of the signal. 

\fakeparagraph{\acf{CF}}
\ac{CF} is the centre frequency, which can be programmed in steps of \SI{61}{\Hz} between \SI{137}{\MHz} to \SI{1020}{\MHz}. 

\fakeparagraph{\acf{SF}}
\ac{SF} determines how many bits are encoded in each symbol, and can be set between 6 and 12. A higher spreading factor increases the \ac{SNR} and therefore receiver sensitivity and range of the signal. However, it lowers the transmission rate and thus increases the transmission duration and energy consumption. The \ac{SF}s in LoRa are orthogonal. Consequently, concurrent transmissions with different SF do not interfere with each other, and can be successfully decoded (assuming a receiver with multiple receive paths).

\fakeparagraph{\acf{BW}}
\ac{BW} can be set from (a fairly narrow) \SI{7.8}{\kHz} up to \SI{500}{\kHz}. In a typical LoRa deployment, only \SI{125}{\kHz}, \SI{250}{\kHz} and \SI{500}{\kHz} are considered. A wider bandwidth means a more spread-out and therefore more interference-resilient link. In addition, it increases the data rate, as the chips are sent out at a rate equivalent to the bandwidth. The downside of a higher bandwidth is a less sensitive reception, caused by the integration of additional noise.

\fakeparagraph{\acf{CR}}
\ac{CR} is the amount of \ac{FEC} that is applied to the message to protect it against burst interference. Higher \ac{CR} makes the message longer and therefore increases the time on air. LoRa transceivers with different \ac{CR}, and operating in `explicit header mode', can still communicate with each other, as the \ac{CR} is encoded in the header.

\section{\ac{LoRa} Simulation Environment}\label{sec:sim}

In our previous work~\cite{bor16lora} we investigated the general scalability of \ac{LoRa} networks. For this study we carried out  testbed experiments to characterise \ac{LoRa} link behaviour. We then used the results of this study to develop the simulation tool \simulator{}\footnote{Available at http://www.lancaster.ac.uk/scc/sites/lora/}.
\simulator models (i) achievable communication range in dependence of communication settings TP, SF and BW and (ii) capture effect behaviour of LoRa transmissions depending on transmission timings and power. We extend \simulator  for the experiments in this paper with (iii) the ability to simulate directional transmissions. Correct representation of these three effects is important as they determine if interfering transmissions can be decoded by a receiver. How effect (i) and (ii) are represented by \simulator is described in detail in our previous publication~\cite{bor16lora}; we include here a brief summary. 
 
\fakeparagraph{\simulator}
\simulator is a custom-build discrete-event simulator implement with SimPy~\cite{simpy}. \simulator allows us to place $N$ LoRa nodes and $M$ LoRa base stations in a 2-dimensional space. The communication characteristics  of a \ac{LoRa} node are defined by the transmission parameters \ac{TP}, \ac{CF}, \ac{SF}, \ac{BW} and \ac{CR}. Furthermore, a node's transmission behaviour is described by the average packet transmission rate $\lambda$ and the size of the packet payload $B$. 

\simulator emulates \ac{LoRa} base station chips such as the Semtech SX1301. This chip can receive up to eight concurrent signals as long as these signals are orthogonal, that is, they use different \ac{SF}.

\fakeparagraph{Communication Range}
A transmission is successfully received if the received signal power $P_{rx}$ lies above the sensitivity threshold $S_{rx}$ of the receiver.  The  received signal power $P_{rx}$  depends on the transmit power $P_{tx}$ and all gains and losses along the communication path. We use the well known log-distance path loss model~\cite{rappaport1996wireless} which is commonly used to model deployments in built-up and densely populated areas. 

On the transmitter side, range can only be changed by changing the transmit power. The range can also be influenced by the use of a directional antenna (described later).  Other parameters like \ac{SF}, \ac{BW} and \ac{CR} do not influence the radiated power, or any other gains and losses. On the receiver side, the range is limited by the sensitivity threshold $S_{rx}$, which is influenced by the \ac{LoRa} parameters \ac{SF} and \ac{BW}. 

To determine $P_{rx}$, the  path loss model must be configured and the communication distance $d$ must be known. In our simulations we configure the path loss to reflect a built up environment. $S_{rx}$ depends on the selected \ac{BW} and \ac{SF}. We use the measured sensitivity from calibration experiments  based on the Semtech SX1272 LoRa transceiver to determine sensitivity in dependence of  \ac{BW} and \ac{SF}. 

\fakeparagraph{Collision Behaviour}
When two \ac{LoRa} transmissions overlap at the receiver, there are several conditions which determine whether the receiver can decode one or two packets, or nothing at all. These conditions are \acf{CF}, \acf{SF}, power and timing. As \ac{LoRa} is a form of frequency modulation, it exhibits the \emph{capture effect}. The capture effect occurs when two signals are present at the receiver and the weaker signal is suppressed by the stronger signal. The difference in received signal strength can therefore be relatively small. An increase of signal strength as present when using a directional antenna has significant impact on this behaviour.

Collision behaviour including capture effect is modelled in \simulator to match  a Semtech SX1272. 

\fakeparagraph{Directional Antenna}
We extend \simulator with directional transmissions. We model  our transmissions according to the \spida antenna~\cite{nilsson2010spida}, an electronically switchable directional (ESD) antenna designed for low-power wireless sensor networks. \spida has six parasitic elements that can be individually grounded or isolated via a software control at negligible energy cost. If all the parasitic elements except one are grounded, the direction of the maximum antenna gain is towards the isolated element. In the experiments we let the direction of maximum gain point towards the receiving base station. As a result, the received signal power of transmissions at the intended receiver increases while it might increase or decrease at other receivers depending on their location. This increase or decrease is based on our previous measurements with \spida~\cite{varshney2013directional,voigt2013understanding}. We emulate an antenna that behaves approximately as \spida with a gain of \SI{4}{\dBi} in the main direction, i.e., when the parasitic element pointing towards the base station is isolated. When the two neighbouring elements are isolated, the same gain is achieved. If the parasitic element opposed the base station is isolated, the gain is decreased with \SI{3}{\dBi}, while for the other two elements the gain is decreased with \SI{4}{\dBi}. We also emulate an improved theoretical antenna where we double these values. For example, in the direction towards the base station, the gain with this antenna is \SI{8}{\dBi}.

\section{Evaluation}\label{sec:eval}

We use  \ac{DER} as metric for evaluating network performance. \ac{DER} is defined as the ratio of received messages to transmitted messages over a  period 
of time. Note that a message is regarded as received correctly if at least one \ac{LoRa} base station of the corresponding network receives it. \ac{DER}  does not capture individual node performance but looks at the network deployment as a whole. When all transmitted messages arrive successfully at one of the base stations, then $DER=1$.

All experiments use the same node configuration set,  $SN = \{TP, CF, SF, BW, CR, \lambda, B\}$. In particular, we study a set we call $SN^1$ where   $TP=\SI{14}{\dBm}$, $CF=\SI{868}{\MHz}$, $SF=12$, $BW=\SI{125}{\kHz}$, $CR=4/8$, $\lambda=\SI{16.7}{min}$ and $B=\SI{20}{\byte}$. $SN^1$ corresponds to the most robust \ac{LoRa} transmitter settings. $SN^1$ transmissions have the longest possible airtime: \SI{1712.13}{\milli\second}.  Due to space constraints, we do not present results for other node settings. We have, however, verified in our simulations that other settings and in particular a setting called $SN^3$~\cite{bor16lora} shows the same trends. $SN^3$ is similar to $SN^1$ except for a lower coding rate which reduces the time on air and leads to fewer collisions. Current \ac{LoRa} deployments use static configurations such as  $SN^1$ or $SN^3$; for example, \ac{LoRaWAN} based deployments use $SN^3$. 

\begin{figure}[t]
  \centering
  \includegraphics[width=0.9\columnwidth]{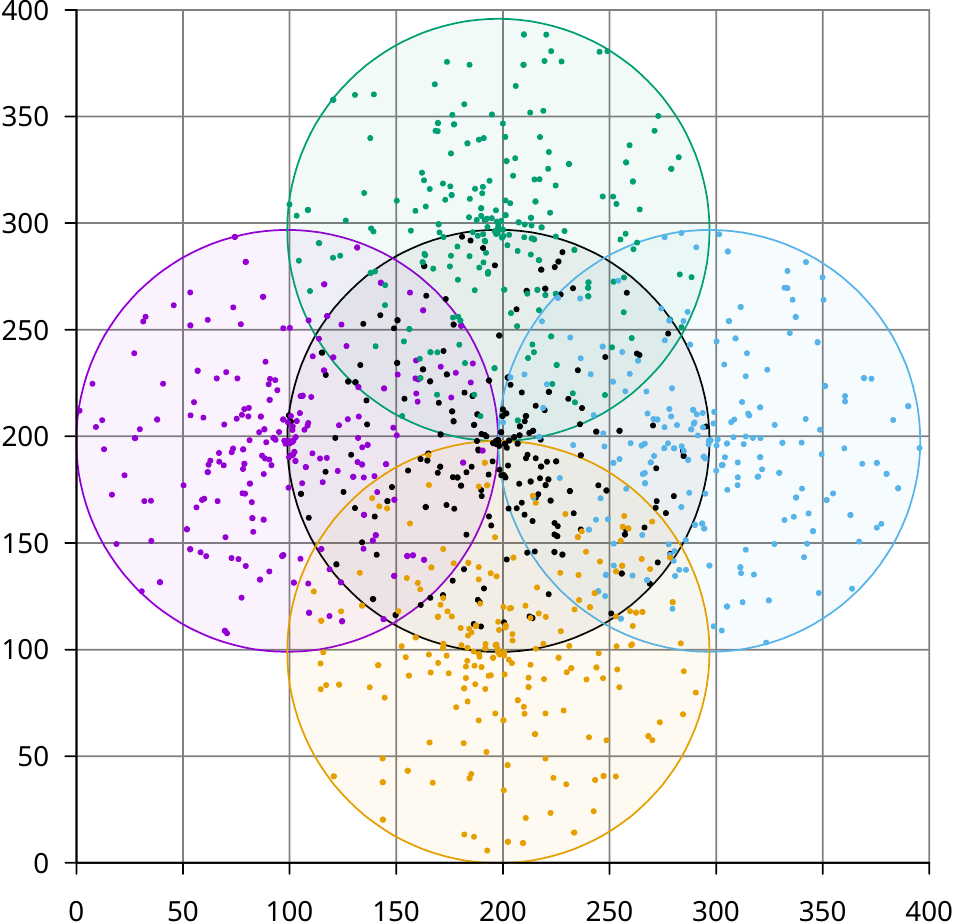}
  \caption{Example configuration used in our simulations. The black network in the centre is interfered by four other networks.}\label{fig:setup}
\end{figure}

In our experiments we create networks by placing $N$ nodes randomly within a circle of radius $R$ around a base station. The distance between nodes and base stations is such that all nodes can reach the base station with the given transmitter settings. If no interference occurs transmissions of nodes reach their base station without loss. In the experiments we use a radius of $R= \SI{99}{\metre}$ which represents a realistic range for built-up environments~\cite{bor2016lora}.  In the experiments we deploy a variable number of interfering networks around the network of interest (called the interfered network); interfering networks are deployed in the same way as our main network. Figure~\ref{fig:setup} shows an example configuration; the black network is the network of interest; the other 4 networks are interfering systems. In all experiments, we assume a 20~Byte packet is sent by each node every \SI{16.7}{\minute} representing a realistic application; the main network and interfering networks use this transmission pattern.  

\subsection{Impact of Interfering Networks}
We use a setup as depicted in Figure~\ref{fig:setup} and described previously. While transmitting to the base station transmissions might interfere with transmissions from other networks (or with nodes from their own network) depending on their position. The network of interest is the one in the centre (see Figure~\ref{fig:setup}). Our goal is to evaluate performance of this network in form of \ac{DER}. In the first experiment, we assume that each network has $N = 200$ nodes. We vary the distance between the base stations of the interfering networks to the base station of the interfered network. The purpose of this first experiment is to show how inter-network interference impacts on \ac{DER}. In the second experiment, we vary the number of nodes per network and the number of interfering networks. The distance between the base stations in this second experiment is \SI{99}{\metre} which is also the network radius. 

\begin{figure}[t]
  \centering
  \includegraphics[width=\columnwidth]{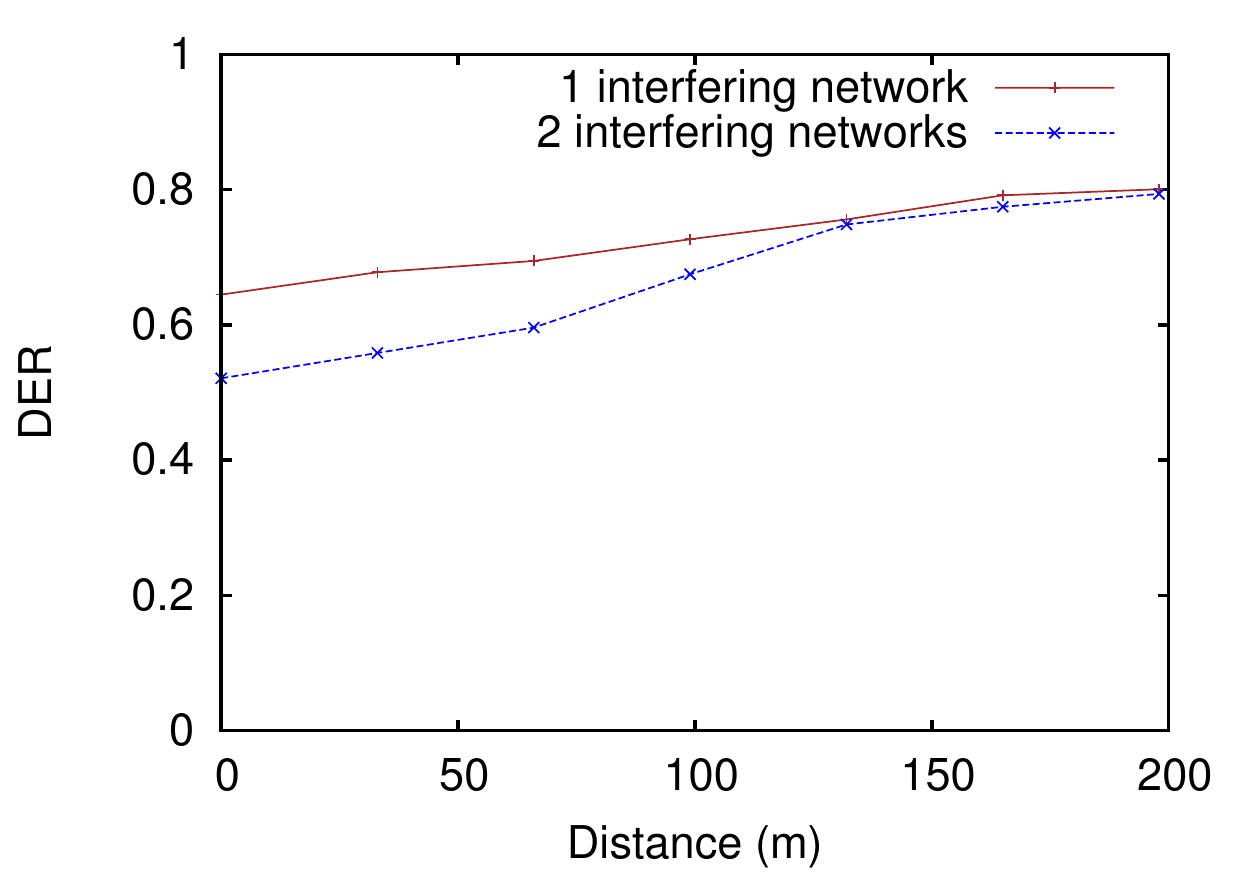}
  \caption{When the distance to the interfering base stations increases, the \ac{DER} of a deployed LoRa network increases as expected.}\label{fig:distBER}
\end{figure}

The results of the first experiment are shown in Figure~\ref{fig:distBER}. When all base stations are placed at the same location, \ie{} the distance is zero, the interference is the highest since the transmissions of all nodes in the interfering network can interfere with the transmissions of the interfered network. With an increasing distance, less nodes of the interfering networks interfere with the transmissions of the interfered networks which leads to a higher \ac{DER}. When the distance between the base stations is \SI{200}{\metre}, no interference between the base
stations is possible. \ac{DER} is here around 0.8 which is the maximum achievable performance due to interference from within the own network.

\begin{figure}[t]
  \centering
  \includegraphics[width=\columnwidth]{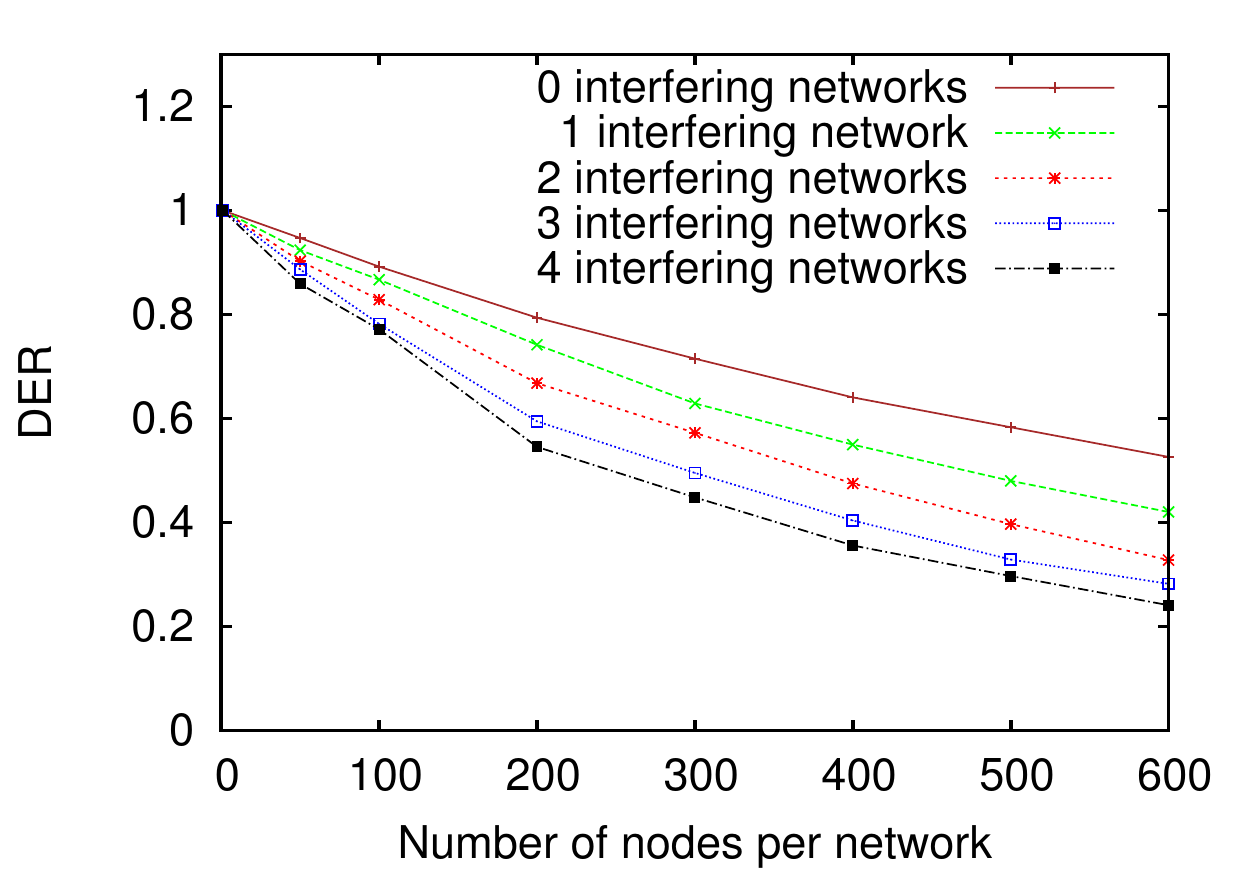}
  \caption{With more interfering networks, \ac{DER} decreases significantly in particular when the number of nodes is high.}\label{fig:nrBS-DER}
\end{figure}

Figure~\ref{fig:nrBS-DER} depicts the results of the second experiment. As expected, when the number of interfering networks increases,  \ac{DER} decreases significantly, in particular when the number of nodes is high. For example, with 500 nodes per network, the \ac{DER} of the interfered network decreases from 0.58 without interference to ca.\@ 0.3 when there is interference from four networks.

The experiments show that the deployment of non-cooperating \ac{LoRa} networks in the same space has a significant impact on network performance. Hence, it is desirable to address and mitigate this issue.

\subsection{Using Directional Antennae}
In the experiments in this section we evaluate to which extend directional antennae can improve the \ac{DER} of an interfered \ac{LoRa} network. We expect that this is possible as the directional antennae can radiate more energy towards the intended base station thereby increasing the signal strength at the base station. Directional antennae also reduce the interference at other base stations which is likely to increase the overall performance of all networks. This is, however, not the focus of this study. The experimental setup is equivalent to the previously used setups; however, nodes in the network of interest (the centre network in Figure~\ref{fig:setup}) are equipped with directional antennae. 

\begin{figure}[t]
  \centering
  \includegraphics[width=\columnwidth]{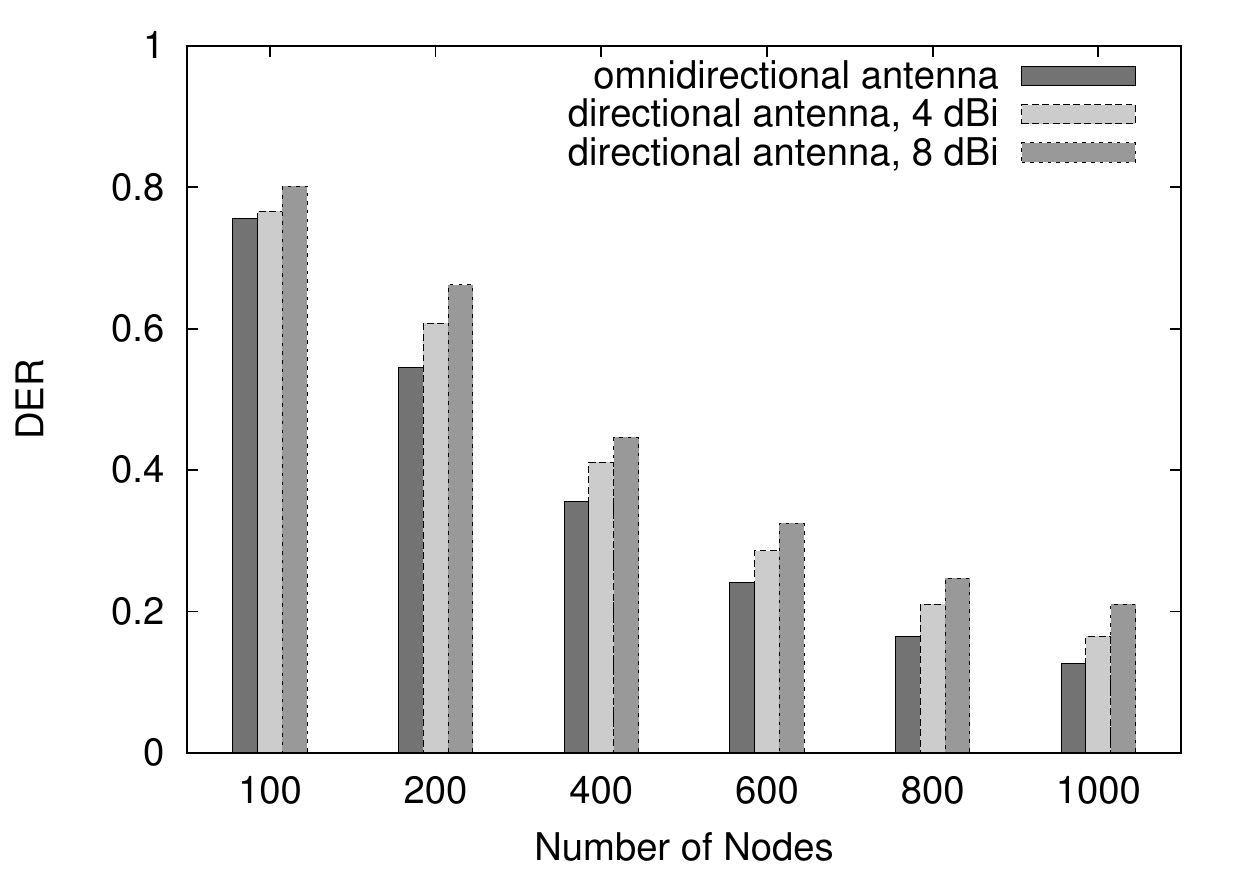}
  \caption{Comparison with omnidirectional antenna, directional antenna and better directional antenna (8~dBi) for four interfering networks. Directional antennae increase the DER.}
\label{fig:newComp}
\end{figure}

Figure~\ref{fig:newComp} depicts the results. The figure shows that as expected, directional transmissions improve \ac{DER}, in particular when the number of nodes is high. This is the case
as the signal strength of  nodes of the interfered network increases. As consequence, it is more likely that due to the capture effect these transmissions succeed even when there are collisions. For most of the setups, \ac{DER} increases by about 0.04 when we equip the nodes in the interfered network  with directional antennae. Using even better directional antennae (8dBi gain), the \ac{DER} increases by another 0.04. 

\subsection{Using Additional  Base Stations}
Our previous work has shown that one way to make \ac{LoRa} networks scale is to increase the number of base stations~\cite{bor16lora}. In the experiments in this section, we evaluate whether this is also true in interference settings. 

We replace the base station in the centre of the setup shown in Figure~\ref{fig:setup} by two and three base stations respectively. We place these additional base stations at a distance $d$ from the original base station. For two stations  we move the original base station $d$  to the right (leaving its vertical position as it is) and add an additional base station $d$ to the left of the original location. When replacing the original base station with three base stations, we move one base station upwards by $d$ and the other two \ang{45} down and to the left and right respectively, so that the distance is also $d$ from the original location of the base station. The placement of the sensor nodes is unchanged, \ie{} they are placed within the radius $r$ around the original location of the base station.  A packet transmission is counted as successful if either of the base stations receives it. All four interfering networks are active.

\begin{figure}[t]
  \centering
  \includegraphics[width=\columnwidth]{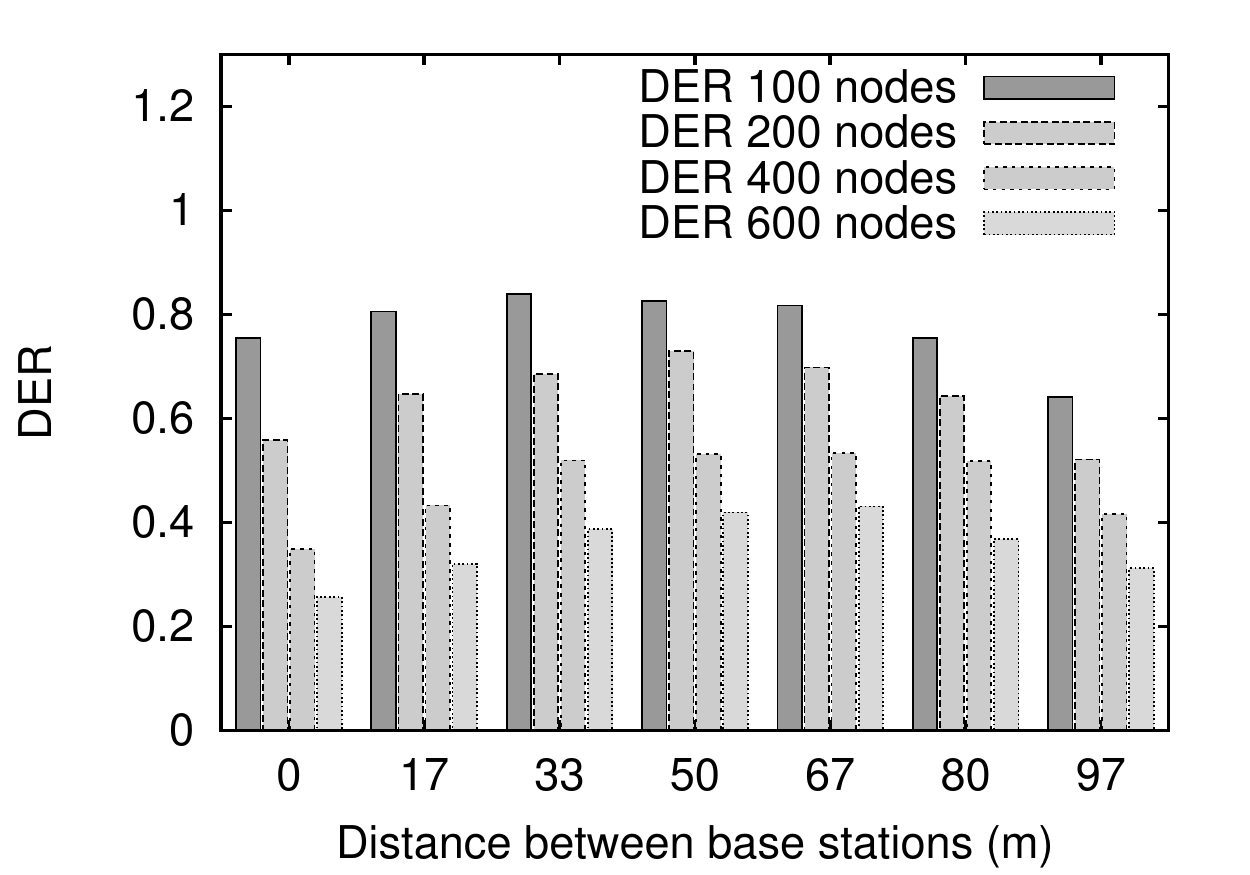}
  \caption{Impact of distance between two base stations on \ac{DER} for different number of nodes.}\label{fig:moreBS2}
\end{figure}

Figure~\ref{fig:moreBS2} shows the results when the original base station is replaced with two base stations.  The figure depicts that with a distance of \SI{0}{\metre} the \ac{DER} is quite low. There is no improvement compared to the results with the omnidirectional antennae in Figure~\ref{fig:newComp}: placing two base stations at the same place does not change anything as they will receive exactly the same packets. For larger distances like \SI{97}{\metre}, the \ac{DER} is even lower since some nodes might not even reach the base station. The best distances are between \SI{33}{\metre} and \SI{67}{\metre} for all setups. For 600 nodes, the \ac{DER} increases from 0.35 to 0.53 for distances between the base stations of \SI{50}{\metre} and \SI{67}{\metre}. 

\begin{figure}[t]
  \centering
  \includegraphics[width=\columnwidth]{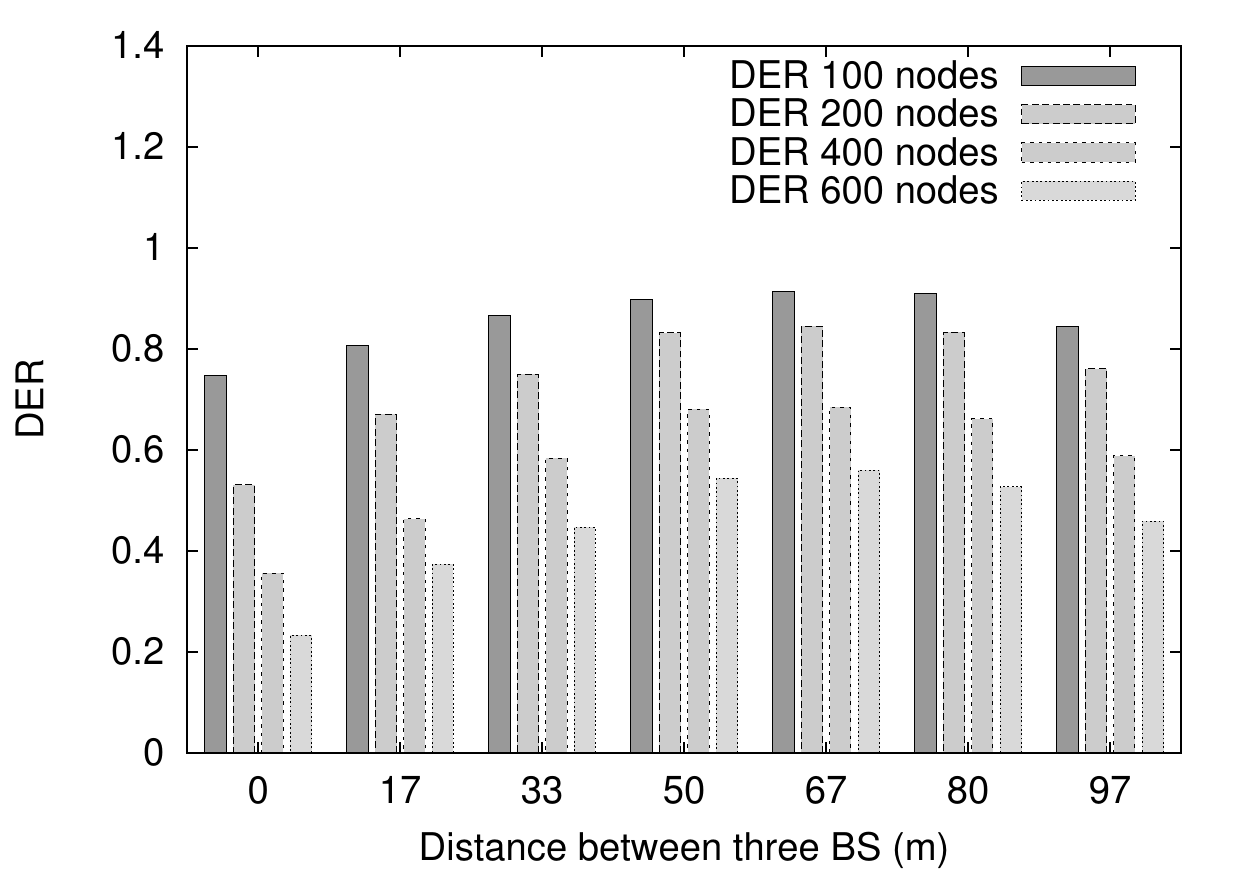}
  \caption{Impact of distance between three base stations on \ac{DER} for different number of nodes.}\label{fig:moreBS3}
\end{figure}

Figure~\ref{fig:moreBS3} depicts the results when the original base station is replaced with three base stations. In general, while the trends are similar to those in Figure~\ref{fig:moreBS2}, the \ac{DER} is higher than with two base stations. In particular, the results with larger distance, \eg{} \SI{97}{\metre} are much better. The reason is the distribution of the base station that ensures that all nodes are in reach of a base station which was not the case for two base stations. Also, the overall results are higher since the chance that a transmission finds a base station where the capture effect comes into play increases. 

\subsection{Discussion}

\begin{figure}[t]
  \centering
  \includegraphics[width=\columnwidth]{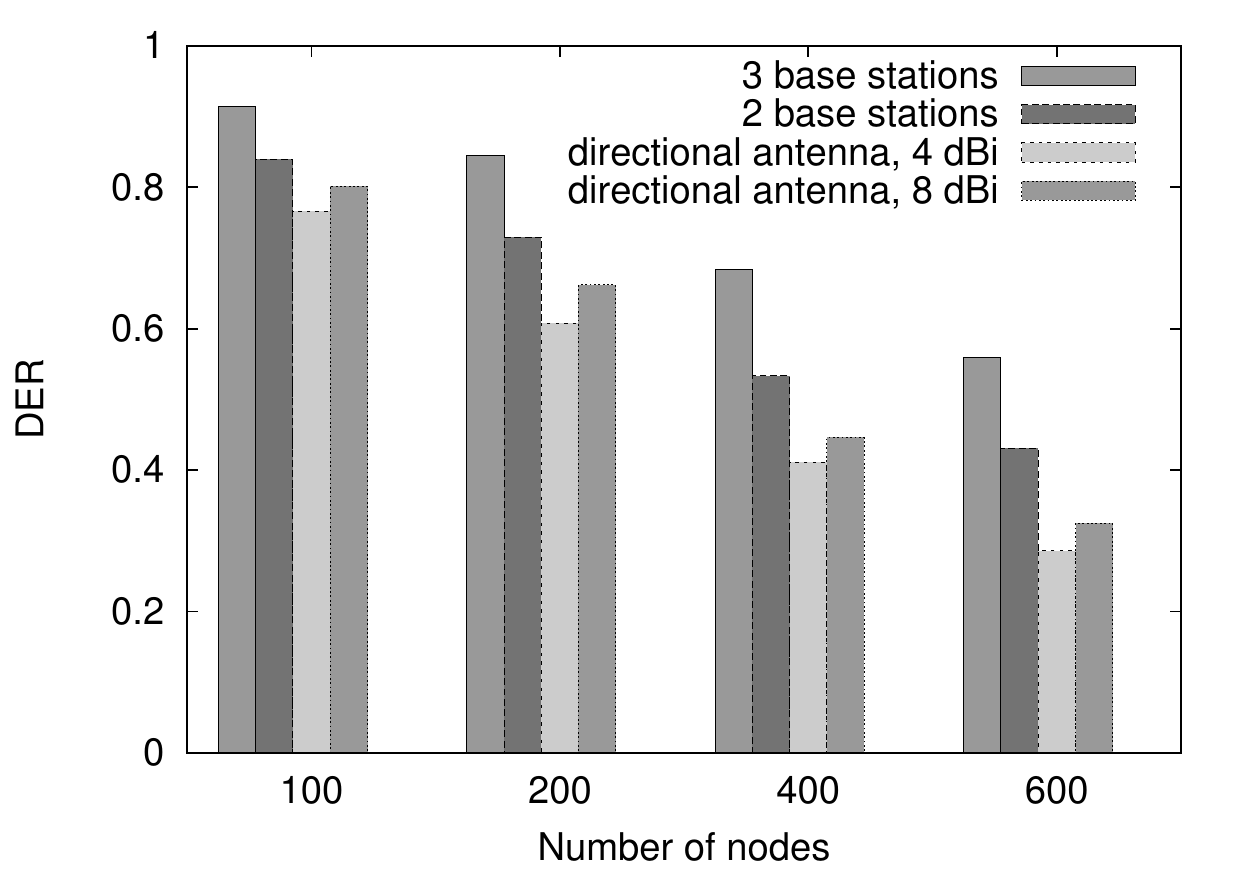}
  \caption{Summary of results: Deploying multiple base stations is more efficient than using directional antennae.}\label{fig:orDir}
\end{figure}

Using the experimental results, we can now answer the question if it is better to equip sensor nodes with directional antennae or to deploy additional base stations to achieve a high \ac{DER} under interference. The result in Figure~\ref{fig:orDir} shows that to achieve a high \ac{DER} under interference, deploying multiple base stations is more efficient than using directional antennae. Moreover, Figure~\ref{fig:nrBS-DER} shows that with multiple base stations, \ac{LoRa} can achieve a DER that is higher than for one base station without inter-network interference. Even when there is no inter-network interference, the transmissions of the nodes from the own network can cause collisions. In our previous study without interference, we have already seen that  using multiple base stations is an efficient way to scale \ac{LoRa} networks~\cite{bor16lora}. Note that from a practical point of view, it is also easier to deploy multiple base stations than to equip sensor nodes with directional antennae, in particular for a  sub-1~\si{\GHz} frequency where antennae are larger in size than antennae  for higher frequency bands. Combining both methods (additional base stations and directional antennae) is theoretically possible but seems an impractical choice.

\section{Related Work}\label{sec:related_work}

There is limited published work discussing interference issues and scalability of \ac{LoRa}. Closest to this  paper is the work by \citeauthor{petajajarvi15coverage} who present an  evaluation of \ac{LoRa} link behaviour in open spaces~\cite{petajajarvi15coverage}.
In another paper~\cite{petyjyjyrvi2016evaluation}, the same authors evaluate the coverage and reliability of a \ac{LoRa} node operating close to a human in an indoor area. The authors also analyse the capacity and scalability of \ac{LoRa} in a more general approach~\cite{mikhaylov2016analysis}  but in contrast to our work~\cite{bor16lora} this seems to be based mostly on the theoretical data rather than real-world calibrated simulations as we do~\cite{bor2016lora}. 
\citet{georgiou2016low} show that the performance drops exponentially as the number of end-devices grows, similar to what we have see in our previous work~\cite{bor16lora}. None of these previous efforts considers the case of interference from co-located, non-cooperating \ac{LoRa} networks.

\citeauthor{saifullah2016snow} present SNOW~\cite{saifullah2016snow}, a long-range
sensor network that operates on white spaces and in many aspects is similar
to \ac{LoRa}. They present a distributed implementation
of OFDM that allows them to decode a large number of
concurrent transmissions. In contrast, we assume a base station that
has more constraints which limits scalability.

Directional antennae have been widely used in cellular and other
wireless networks~\cite{dai2013overview}.
In the context of wireless sensor networks, \citeauthor{mottola13electronically} show how only minor 
modifications to an existing protocol originally designed 
for omnidirectional antennae can bring performance improvements
when using directional antennae~\cite{mottola13electronically}. \citeauthor{varshney2015directional} use them to improve the performance of bulk transfers~\cite{varshney2015directional}.

\section{Conclusions}

In this paper we have evaluated the impact of inter-network interference on \ac{LoRa} networks. Through simulations based on real experimental data, we have shown that interference can drastically reduce the performance of a \ac{LoRa} network.
Our results demonstrate that directional antennae and using multiple base stations can improve performance under interference. Our simulations show that 
deploying multiple base stations outperforms the use of directional antennae.


\balance

\bibliographystyle{abbrvnat}
\bibliography{sigproc} 

\begin{thebibliography}{15}
\providecommand{\natexlab}[1]{#1}
\providecommand{\url}[1]{\texttt{#1}}
\expandafter\ifx\csname urlstyle\endcsname\relax
  \providecommand{\doi}[1]{doi: #1}\else
  \providecommand{\doi}{doi: \begingroup \urlstyle{rm}\Url}\fi

\bibitem[sim()]{simpy}
{SimPy -- Event discrete simulation for Python}.
\newblock \url{https://simpy.readthedocs.io}.
\newblock Accessed: 24-05-2016.

\bibitem[Bor et~al.(2016{\natexlab{a}})Bor, Roedig, Voigt, and
  Alonso]{bor16lora}
M.~Bor, U.~Roedig, T.~Voigt, and J.~Alonso.
\newblock {Do LoRa Low-Power Wide-Area Networks Scale?}
\newblock In \emph{Proceedings of the Conference on Modeling, Analysis and
  Simulation of Wireless and Mobile Systems (ACM MSWiM)}, 2016{\natexlab{a}}.

\bibitem[Bor et~al.(2016{\natexlab{b}})Bor, Vidler, and Roedig]{bor2016lora}
M.~Bor, J.~Vidler, and U.~Roedig.
\newblock {LoRa for the Internet of Things}.
\newblock In \emph{Proceedings of the 2016 International Conference on Embedded
  Wireless Systems and Networks}, EWSN '16, pages 361--366, USA,
  2016{\natexlab{b}}. Junction Publishing.
\newblock ISBN 978-0-9949886-0-7.

\bibitem[Dai et~al.(2013)Dai, Ng, Li, and Wu]{dai2013overview}
H.~Dai, K.-W. Ng, M.~Li, and M.-Y. Wu.
\newblock An overview of using directional antennas in wireless networks.
\newblock \emph{International Journal of Communication Systems}, 26\penalty0
  (4):\penalty0 413--448, 2013.

\bibitem[Georgiou and Raza(2016)]{georgiou2016low}
O.~Georgiou and U.~Raza.
\newblock Low power wide area network analysis: Can {LoRa} scale?
\newblock \emph{arXiv preprint arXiv:1610.04793}, 2016.

\bibitem[Mikhaylov et~al.(2016)Mikhaylov, Pet\"aj\"aj\"arvi, and
  Haenninen]{mikhaylov2016analysis}
K.~Mikhaylov, J.~Pet\"aj\"aj\"arvi, and T.~Haenninen.
\newblock Analysis of capacity and scalability of the {LoRa} low power wide
  area network technology.
\newblock In \emph{22th European Wireless Conference}, pages 1--6, 2016.

\bibitem[Mottola et~al.(2013)Mottola, Voigt, and
  Picco]{mottola13electronically}
L.~Mottola, T.~Voigt, and G.~Picco.
\newblock Electronically-switched directional antennas for wireless sensor
  networks: A full-stack evaluation.
\newblock In \emph{The IEEE Communications Society Conference on Sensor, Mesh
  and Ad Hoc Communications and Networks (IEEE SECON)}, 2013.

\bibitem[Nilsson(2010)]{nilsson2010spida}
M.~Nilsson.
\newblock Spida: A direction-finding antenna for wireless sensor networks.
\newblock In \emph{Organization of the Workshop on Real-World Wireless Sensor
  Networks (REALWSN)}, pages 138--145. Springer, 2010.

\bibitem[Pet\"aj\"aj\"arvi et~al.(2015)Pet\"aj\"aj\"arvi, Mikhaylov, Roivainen,
  Hanninen, and Pettissalo]{petajajarvi15coverage}
J.~Pet\"aj\"aj\"arvi, K.~Mikhaylov, A.~Roivainen, T.~Hanninen, and
  M.~Pettissalo.
\newblock On the coverage of {LPWANs}: range evaluation and channel attenuation
  model for {LoRa} technology.
\newblock In \emph{ITS Telecommunications (ITST), 2015 14th International
  Conference on}, pages 55--59, Dec 2015.
\newblock \doi{10.1109/ITST.2015.7377400}.

\bibitem[Pet\"aj\"aj\"arvi et~al.(2016)Pet\"aj\"aj\"arvi, Mikhaylov,
  H\"am\"al\"ainen, and Iinatti]{petyjyjyrvi2016evaluation}
J.~Pet\"aj\"aj\"arvi, K.~Mikhaylov, M.~H\"am\"al\"ainen, and J.~Iinatti.
\newblock Evaluation of {LoRa LPWAN} technology for remote health and wellbeing
  monitoring.
\newblock In \emph{2016 10th International Symposium on Medical Information and
  Communication Technology (ISMICT)}, pages 1--5. IEEE, 2016.

\bibitem[Rappaport et~al.(1996)]{rappaport1996wireless}
T.~S. Rappaport et~al.
\newblock \emph{Wireless communications: principles and practice}, volume~2.
\newblock Prentice Hall PTR New Jersey, 1996.

\bibitem[Saifullah et~al.(2016)Saifullah, Rahman, Ismail, Lu, Chandra, and
  Liu]{saifullah2016snow}
A.~Saifullah, M.~Rahman, D.~Ismail, C.~Lu, R.~Chandra, and J.~Liu.
\newblock {SNOW}: Sensor network over white spaces.
\newblock In \emph{Proceedings of the International Conference on Embedded
  Networked Sensor Systems (ACM SenSys)}, 2016.

\bibitem[Varshney et~al.(2013)Varshney, Voigt, and
  Mottola]{varshney2013directional}
A.~Varshney, T.~Voigt, and L.~Mottola.
\newblock Directional transmissions and receptions for high throughput burst
  forwarding.
\newblock In \emph{Proceedings of the International Conference on Embedded
  Networked Sensor Systems (ACM SenSys)}. ACM, 2013.

\bibitem[Varshney et~al.(2015)Varshney, Mottola, Carlsson, and
  Voigt]{varshney2015directional}
A.~Varshney, L.~Mottola, M.~Carlsson, and T.~Voigt.
\newblock Directional transmissions and receptions for high-throughput bulk
  forwarding in wireless sensor networks.
\newblock In \emph{Proceedings of the 13th ACM Conference on Embedded Networked
  Sensor Systems}, 2015.

\bibitem[Voigt et~al.(2013)Voigt, Mottola, and Hewage]{voigt2013understanding}
T.~Voigt, L.~Mottola, and K.~Hewage.
\newblock Understanding link dynamics in wireless sensor networks with
  dynamically steerable directional antennas.
\newblock In \emph{European Conference on Wireless Sensor Networks}, pages
  115--130. Springer, 2013.

\end{thebibliography}


\end{document}